\documentclass{iopart}

\usepackage[dvipdfm]{hyperref}
\usepackage[dvipdfm]{graphicx}
\usepackage{url}
\usepackage{bm}
\expandafter\let\csname equation*\endcsname\relax
\expandafter\let\csname endequation*\endcsname\relax
\usepackage{amsmath}

\begin{document}

\title[Simulated tempering and magnetizing of Potts model]{Application of Simulated Tempering and Magnetizing to a Two-Dimensional Potts Model}

\author{Tetsuro NAGAI$^1$, Yuko OKAMOTO$^{1,2,3,4}$, and Wolfhard JANKE$^{5}$}
\address{$^1$Department of Physics, Graduate School of Science, Nagoya University, Nagoya, Aichi 464-8602, Japan} 
\address{$^2$Structural Biology Research Center, Graduate School of Science, Nagoya University, Nagoya, Aichi 464-8602, Japan}
\address{$^3$Center for Computational Science, Graduate School of Engineering, Nagoya University, Nagoya, Aichi 464-8603, Japan}
\address{$^4$Information Technology Center, Nagoya University, Nagoya, Aichi 464-8601, Japan}
\address{$^5$Institut f\"ur Theoretische Physik and Centre for Theoretical Sciences (NTZ), Universit\"at Leipzig, Postfach 100 920, 04009 Leipzig, Germany}

\date{\today}

\begin{abstract}
We applied  the simulated tempering and magnetizing (STM) method to the two-dimensional three-state Potts model 
in an external magnetic field
in order to perform further investigations of the STM's applicability.
The temperature as well as the external field are treated as dynamical variables updated during the STM simulations.  
After we obtained adequate information for several lattice sizes 
$L$ (up to $160\times 160$), 
we also performed a number of conventional canonical simulations of large lattices, 
especially in order to illustrate the crossover behavior of the Potts model in external field with increasing  $L$.
The temperature and external field for larger lattice size simulations were chosen by extrapolation of the detail information obtained by STM.
We carefully analyzed the crossover scaling at the phase transitions with respect to the lattice size as well as  the temperature and external field. 
The crossover behavior is clearly observed in the simulations in agreement with theoretical predictions.

\end{abstract}


\maketitle

\section{Introduction}

Monte Carlo (MC) and molecular dynamics (MD)  methods have been demonstrated in many applications to be indispensable 
tools to study 
the statistical properties of various systems in equilibrium.
The quasi-ergodicity problem, however, where the system gets trapped 
in states of local energy minima, has often posed great difficulties. 
In order to overcome this difficulty, generalized-ensemble algorithms 
have been developed and applied to many problems 
including spin models and biomolecular systems 
(for reviews, see, e.g., Refs.\ \cite{Janke1998, Hansmann1999,Mitsutake2001,janke2008rugged}).

Well-known examples of generalized-ensemble algorithms are  the multicanonical algorithm (MUCA) 
\cite{berg1991multicanonical,berg1992multicanonical}, 
simulated tempering (ST)  \cite{Lyubartsev1992,marinari1992simulated}, 
and the replica-exchange method (REM) \cite{hukushima1996exchange,Geyer1991} (also referred to as parallel tempering). 
Closely related to MUCA are the Wang-Landau method \cite{Wang2001a,Wang2001b} and metadynamics \cite{Laio2002}.
REM is implicitly a special case of the general method described in the earlier work of Ref.\ \cite{Swendsen1986}, as
detailed later in Ref.~\cite{wang2004replica}. 

Based on the recent multi-dimensional generalization of  generalized-ensemble algorithms \cite{Mitsutake2009multidimensional1,Mitsutake2009multidimensional2,Mitsutake2009MSTMREM},   
the ``Simulated Tempering and Magnetizing'' (STM) method has been proposed and developed \cite{Nagai2012Proc,Nagai2012inpress}.  
In Refs.~\cite{Nagai2012Proc, Nagai2012inpress} two of us studied the classical Ising model, 
introducing the external (magnetic) field as a second dynamical variable besides the temperature
and showed improvements over the conventional ``one-dimensional'' simulated tempering schemes, such as 
better sampling efficiency and potential applicability to a first-order phase transition, which cannot be dealt with by one-dimensional ST.

In the present work, we further investigate the STM method, applying it to the two-dimensional three-state Potts model 
in external magnetic field \cite{Potts1952,Wu1982}. 
This model has several interesting applications in condensed matter physics \cite{Wu1982} and its three-dimensional 
counterpart serves as an effective model for quantum chromodynamics \cite{philipsen2006qcd,kim20053,Karsch2000,Mercado2012}.
We see the STM scheme working in this more complicated system as well. 
We also look into crossover behaviors according to lattice size $L$ as well as temperature $T$ and external field $h$. 
We observe that the STM method lets us investigate a wide area of sampling space. 

The rest of this article is organized as follows. In section 2 we review the STM method and give the details of our simulations. 
In section 3 we present the results. After checking the two-dimensional random walks, we compare ST and STM, and 
calculate various thermodynamic quantities at many sets of parameter values, in combination with reweighting techniques. 
We then use this extensive data set to study the crossover scaling behavior at the phase transitions with respect to the
lattice size as well as the temperature and external field.
In section 4 we conclude this article.

\section{Model and Methods}

\subsection{Model}
We study the two-dimensional three-state standard Potts model in external field with energy 
\begin{align}
H&=E-hM\,, \label{eq:H}\\
E&=-\sum_{\left< i,j\right>} \delta_{\sigma_i, \sigma_j} \label{eq:E}\,,\\
M&=\sum_{i=1}^N \delta_{0, \sigma_i} \,,\label{eq:M}
\end{align}
where $N=L^2$ denotes the total number of spins, $\delta$ is the Kronecker delta function, 
$\sigma_i$ a spin at the $i$-th site, and $h$ the external field.
The spin $\sigma_i$ takes on one of the three values $0$, $1$, or $2$.
The sum in (\ref{eq:E}) goes over all nearest-neighbor pairs, where the spins are arranged on a square $L\times L$ lattice
with periodic boundary conditions. 
Data were obtained in STM for lattice sizes ranging from $2\times2$ to $160\times 160$
and additionally in conventional canonical simulations on $320\times 320$ and $640\times 640$ lattices.

We recall that the three-state (standard)  Potts model is equivalent to the three-state planar Potts or $Z_3$ model.
We first introduce a spin 
\begin{align}
\vec{s}_i &= \left( \begin{matrix}\cos \frac{2 \pi}{3} \sigma_i \\  \sin \frac{2 \pi}{3} \sigma_i \end{matrix} \right). \label{eq:vec} 
\end{align} 
The zero-field energy term is then given by 
\begin{align}
E_{\mathrm{planar}}=-\sum_{\left<i,\,j \right>} \vec{s}_i\cdot\vec{s}_j\,,
\end{align} 
and the magnetization reads 
\begin{align}
\vec{M}=\sum_{i=1}^{N} \vec{s}_i.
\end{align} 
In an external field chosen along the $x$-direction,  
\begin{align}
\vec{h} \equiv h\vec{e}_x,
\end{align} 
the product of external field and magnetization becomes 
\begin{align}
\vec{h}\cdot\vec{M}=h M^{(x)},
\end{align} 
where $M^{(x)}$ stands for the $x$-component of $\vec{M}$, which is given by $\sum_i \cos \frac{2\pi}{3}\sigma_i$.  
Because 
the argument of the cosine and sine in (\ref{eq:vec}) can only take the values of $0$, $2\pi/3$, and $4\pi/3$, we have  
\begin{align}
E&=\left( E_{\mathrm{planar}} - N \right)\times \frac{2}{3} \,,\\ 
M&=\left( M^{(x)}+\frac{N}{2} \right)\times \frac{2}{3} \,. 
\end{align}
Thus, we arrive at  
\begin{align}
E-hM&=\frac{2}{3}E_{\mathrm{planar}} -\frac{2}{3}hM^{(x)} -\frac{N}{3}(2-h) \,. 
\end{align}
Because the last term is an unimportant constant, 
the standard Potts model is equivalent to the planar Potts model 
with a $\frac{2}{3}$-smaller coupling constant and
magnetization normalized by $\frac{2}{3}$:  
\begin{align}
H&=E-hM\,, \label{eq:H2}\\
E&=-\frac{2}{3}\sum_{\left< i,j\right>} \cos \theta_{ij} \label{eq:E2}\,,\\
M&=\frac{2}{3}\sum_{i=1}^N \cos \theta_i \,,\label{eq:M2}
\end{align}
where $\theta_{ij}$ and $\theta_{i}$ are defined by $\theta_j -\theta_i $ and $\frac{2\pi}{3} \sigma_i$, respectively.

Because $M^{(x)}$ is the projection of $\vec{M}$ onto the $x$-axis, $M^{(x)}/N$ equals $1$, $-1/2$, $-1/2$, and $0$ when the system is 
in the 0-direction ordered, the 1-direction ordered, the 2-direction ordered, and the disordered state, respectively.   
This corresponds to $M/N=0$ when the system is in  1- or 2-direction ordered states and
$M/N=1/3$ in the disordered state.

\begin{figure}[hbtp]
   \begin{center}
       \includegraphics[width=7cm]{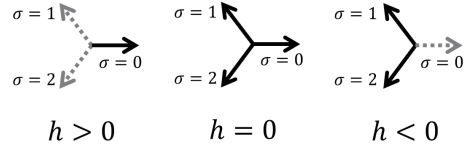}
       \caption{Schematic description of the behavior of spins according to the external field. 
(a) Spin 0 is favored at $h>0$, i.e., spin 1 and spin 2 are disfavored. 
(b) All the three states are equivalent at $h=0$.  
(c) Spin 0 is disfavored at $h<0$. 
}
       \label{fig:spin}
   \end{center}
\end{figure}

As one can see from (\ref{eq:H}) and (\ref{eq:M}) [or, (\ref{eq:H2}) and (\ref{eq:M2})], 
spin direction $0$ is favored by a positive external field ($h>0$).
Accordingly, a negative external field ($h<0$) disfavors spin direction $0$.   
Thus, the system is expected to behave like a two-dimensional Ising model in the presence of a negative external field. 
In fact, in the limit $h\rightarrow -\infty$, the three-state Potts model is equivalent to the Ising model in zero external field, because the 
unfavored states disappear in the partition function calculations.  Figure \ref{fig:spin} illustrates the schematic picture of this relation. 

For the easier availability of reference, we summarize the critical exponents for the two-dimensional Ising 
and three-state Potts model in table \ref{table:exp} \cite{Wu1982}. 
\begin{table}
\caption{Critical exponents for the 
two-dimensional Ising  and three-state Potts models ($y_t=1/\nu$, $y_h=(\beta+\gamma)/\nu$).}
\label{table:exp} 
\lineup
\begin{indented}
\item\begin{tabular}{@{}lcccccc}\br
Model  & $y_t$& $y_h$&   $\beta$ &  $\gamma $&  $\delta$ & $\nu$  \\ \mr
Ising    &   1&15/8& 1/8    &   7/4      &   15       &1\\
Potts  & 6/5&28/15&1/9 &  13/9       &  14       &5/6\\      
\br
\end{tabular}
\end{indented}
\end{table}

\subsection{Simulation methods}
In this section we briefly review the STM method \cite{Nagai2012Proc, Nagai2012inpress}. 
While the conventional ST method \cite{Lyubartsev1992,marinari1992simulated} 
considers the temperature to be the dynamical variable, 
the STM method employs the external field as a second dynamical variable besides temperature.
This algorithm is  
based on the multi-dimensional extension of generalized-ensemble algorithms 
\cite{Mitsutake2009multidimensional1,Mitsutake2009multidimensional2,Mitsutake2009MSTMREM}. 
In other words, we consider 
\begin{align}
\rme^{-(E-hM)/T +a(T,h)} \label{eq:f1}
\end{align}
as a joint probability for $(x,T,h)$ ($\in X \otimes \{ T_1, T_2, \dots, T_{N_T}\} \otimes \{h_1, h_2, \dots h_{N_h}\}$), 
where $a(T,h)$, $x$, and $X$ are a parameter, the (microscopic) state, and the sampling space, respectively. 
Here and hereafter, we set Boltzmann's constant to unity.  
Note that the temperature and external field are discretized into $N_T$ and $N_h$ values, respectively. 

We will find the candidate for $a(T_i,h_j)$ by looking into 
the probability of occupying each set of parameter values.  
It is given by
\begin{align}
P(T_i,h_j) &= 
        \frac{\displaystyle\sum _{k=1} ^{N_T}\displaystyle\sum _{l=1} ^{N_h} 
                       \int \rmd x\, \delta_{ik}\, \delta_{jl}\, \rme^{-\frac{E(x)-h_l M(x)}{T_k} +a(T_k, h_l)} }
             {\displaystyle\sum _{k=1} ^{N_T}\displaystyle\sum _{l=1} ^{N_h} \int \rmd x\,  \rme^{-\frac{E(x)-h_l M(x)}{T_k} +a(T_k, h_l)}}\\
           &=
              \frac{\rme^{-f(T_i,h_j)+a(T_i, h_j)} }
                   {\displaystyle\sum _{k=1} ^{N_T}\displaystyle\sum _{l=1} ^{N_h}  \rme^{-f(T_k,h_l) +a(T_k, h_l)}} \\
           &\propto \rme^{-f(T_i,h_j)+a(T_i, h_j)} \,, \label{eq:P}
\end{align}
where 
\begin{align}
\rme^{-f(T_i,h_j)}=\int \rmd x \rme^{-(E-h_jM)/T_i}. \label{eq:f2}
\end{align}
Thus, the dimensionless free energy $f(T_i,h_j)$ is a proper choice
for $a(T_i,h_j)$ in order to acquire a uniform distribution of the number of samples according to $T$ and $h$.
These values can be estimated by a number of  methods.
For example, one can obtain such values from preliminary simulations and reweighting techniques.   

Any thermal average $\left< A\right> _{T_i,h_j}$ 
at given $T_i$ ($\in \{T_1,T_2,\dots,T_{N_T}\}$) and $h_j$ ($\in \{h_1,h_2,\dots,h_{N_h}\}$) 
can be obtained by the conditional expectation: 
$ \left< A\right>_{T_i, h_j} = \left<A|T_i, h_j\right>_{\mathrm{STM}} $. 
Namely, we have 
\begin{align}
\left< A\right>_{T_i,h_j} = \frac{1}{N_{T_i,h_j}}\sum _{k=1}^{N_{T_i,h_j}} A^{k}_{T_i,h_j} \,,
\end{align}
where $N_{T_i,h_j}$ is the total number of samples obtained at $T_i$ and $h_j$, and $A^{k}_{T_i,h_j}$ represents the $k$-th sample at $T_i$ and $h_j$.

The temperature $T$ or external field $h$ can be updated similarly to the spin $\sigma_i$, 
because they are considered as dynamical variables.
The Metropolis criterion for updating $T$ and $h$ is given by
\begin{align}
w(T_i, h_j &\rightarrow  T_{i'}, h_{j'}) = \min  \left( 1, \frac{P(T_{i'}, h_{j'})}{P(T_i, h_j)}\right)\\
                                   &= \min  \left(1, \exp\left(-\left(\frac{1}{T_{i'}}-\frac{1}{T_i}\right) E+
\left(\frac{h_{j'}}{T_{i'}}-\frac{h_j}{T_i} \right) M + a(T_{i'},h_{j'}) -a(T_i,h_j)\right)\right) \,.
\label{eq:trans}
\end{align}

Once an initial state is prepared, the STM simulations can be performed by repeating the following two steps:   
1.\ We perform a conventional canonical simulation at fixed $T_i$ and $h_j$ for a certain number of MC sweeps. 
2.\ We update the temperature and/or external field by (\ref{eq:trans}) with $a(T,h)=f(T,h)$.

In our implementation, every fixed number of MC sweeps, either $T$ or $h$  was updated 
 (the choice between $T$ and $h$ was made at random)
by (\ref{eq:trans}) to a neighboring value (the choice between two possible neighbors was also made at random). 
Here, one MC sweep consists of $L\times L$ single spin updates.
The number of MC sweeps performed between parameter updates is here  referred to as the parameter-updating period. 

We remark that, as spins can be updated by a number of algorithms, 
other schemes of updating the parameters can be employed \cite{chodera2011replica}. 
There also exists a temperature updating scheme for ST by the Langevin algorithm \cite{zhang2010}.

Table \ref{table} summarizes the  conditions of the present STM simulations. 
According to the previous studies \cite{Nagai2012inpress, sindhikara2008exchange,sindhikara2010},
we updated the parameters frequently. That is, we employed very small parameter-updating periods.  

\begin{table}
\caption{Parameter values for the STM simulations. }
\label{table} 
\lineup
\footnotesize\begin{tabular}{@{}lcccccc}\br
Lattice size   $L$                &         5&         10&              20&        40&        80&         160      \\ \mr
Total MC sweeps            & 220915510&  216277330& 164000000& 164000000&  43605000 &    525000000 \\
Parameter-updating period &        1  &           1&                 1&       1 &         1        &               1\\ 
$T_1, \dots, T_{N_T}$             & $0.1, \dots, 3.5$  & $0.1, \dots, 3.5$ & $0.3, \dots, 3.2$ &$0.3, \dots, 3.2$&$0.5, \dots, 2.0$&$0.5, \dots, 2.0$\\
$h_1, \dots, h_{N_h}$              &$-1.5, \dots, 1.5$&$-1.5, \dots, 1.5$&$-1.5, \dots, 1.5$&$-1.5, \dots, 1.5$&$ -1.5, \dots, 1.5$ &$-1.0, \dots,  0.025$\\
$N_T$                            &      20    &    20     &             40&        40 &       76       &       95\\
$N_h$                             &     21    &    21     &            41 &        41  &       5        1&30\\
$N_{{\rm data}}$ $^{\rm a}$
                                       &     10    &   10    &             10&        10&          10&            10\\ \br
\end{tabular}
$^{\rm a} $ The data were stored every $N_{{\rm data}}$ MC sweeps.
\end{table}

In addition, we also performed conventional canonical simulations. 
Table \ref{table2} lists their details. 
The temperature was chosen by extrapolations of the STM results. 
We estimated the proper temperature by fitting the STM results to $T_\mathrm{max}-T_\mathrm{c}\propto L^{-1/\nu} $, 
where $T_\mathrm{max}$ is the temperature at which the observables take their maxima.  
The Greek letter $\nu$ denotes the critical exponent of the correlation length. 
For vanishing external field we fitted the data to the Potts case ($\nu=5/6$) and in (negative) external field
 to the Ising case ($\nu=1$), respectively.

\begin{table}
\caption{Parameter values for the regular canonical simulations. }
\label{table2} 
\lineup
\footnotesize\begin{tabular}{@{}lcc}\br
Lattice size   $L$            &  320          &   640       \\ \mr
Total MC sweeps
$^{\rm a}$
&     4000000 &     4000000      \\
$(T,\,h)$ $^{\rm b}$
                                  &  (0.995518, 0),  (0.995995, 0),  (0.996490,  0)  & (0.994985, 0), (0.995209, 0), (0.995512,  0)   \\
                                  &  (1.0752077, $-0.5$), (1.077044, $-0.5$), (1.078302, $-0.5$)  & (1.075672, $-0.5$),  (1.076181, $-0.5$), (1.076936, $-0.5$) \\
                                  &  (1.101447, $-1.0$), (1.102321, $-1.0$), (1.103380, $-1.0$)  & (1.100386, $-1.0$), (1.101043, $-1.0$), (1.101811, $-1.0$)  \\
$N_{{\rm data}}$ $^{\rm c}$
                                                                                                                &         10  &  20 \\ \br
\end{tabular}
$^{\rm a}$ The number performed for each set of temperature and external field.\\
$^{\rm b}$ Three temperature values for each external field.\\
$^{\rm c}$ The data were stored every $N_{{\rm data}}$ MC sweeps.\\
\end{table}

As for spin updates, we employed the single-spin update algorithm; we updated spins one by one with the heatbath algorithm.
As for quasi-random-number generator, we used the Mersenne Twister \cite{matsumoto1998mersenne}.

\subsection{Free energy calculations}
The simulated tempering parameters, or free energy, in (\ref{eq:f1}) and (\ref{eq:f2}) can be simply obtained 
by reweighting techniques applied to the results of
preliminary simulation runs 
\cite{Mitsutake2009multidimensional1,Mitsutake2009multidimensional2, Mitsutake2009MSTMREM, mitsutake2000replica}. 
We used two reweighting methods for the free energy calculations. 
One method is the multiple-histogram reweighting method, or Weighted Histogram Analysis Method (WHAM) \cite{Ferrenberg1989,kumar1992weighted,kumar1995multidimensional} 
and the other is the Multistate Bennett Acceptance Ratio estimator (MBAR) 
\cite{shirts2008statistically}, which is based on WHAM. 

The equations of the WHAM algorithm applied to the system are as follows. For details, 
the reader is referred to Refs.\ \cite{Mitsutake2009multidimensional2,kumar1992weighted,kumar1995multidimensional}.
The density of states (DOS) $n(E,M)$ and free energy values $f(T_i,h_j)$ can be obtained from 
\begin{align}
n(E,M) &= \frac{\displaystyle\sum _{i=1}^{N_T}\displaystyle\sum _{j=1}^{N_h} n_{T_i,h_j}(E,M)}
                {\displaystyle\sum _{i=1}^{N_T}\displaystyle\sum _{j=1}^{N_h} N_{T_i,h_j} \exp 
(f(T_i,h_j)-(E-h_jM)/T_i)} \,,  \label{eq:wham1}\\
f(T_i,h_j) &= -\ln \sum_{E,M} n(E,M) \exp (-(E-h_jM)/T_i)   \,, \label{eq:wham2}
\end{align}
where $n_{T_i,h_j}(E,M)$ is the histogram of $E$ and $M$ at $T_i$ and 
$h_j$, and $N_{T_i, h_j}$ is the total number of 
samples obtained at $T_i$ and $h_j$.
By solving these two equations self-consistently by iteration, we can 
obtain $n(E,M)$ and $f(T_i, h_j)$.
The obtained $n(E,M)$ allows one to calculate any thermal average at arbitrary temperature and external field values.
Note that $f(T_i,h_j)$  is determined up to a constant, which sets the zero point of free energy. 
Accordingly, $n(E,M)$ is determined up to a normalization constant.

The MBAR is based on the  following  equations. Namely, by combing (\ref{eq:wham1}) and (\ref{eq:wham2}), the free energy can be written as  
\begin{align}
f(T_i,h_j) = - \ln
\sum_{n=1}^N 
 \frac{ \exp (-(E_n-h_j M_n)/T_i)}
       {\displaystyle\sum_{k=1}^{N_T}\displaystyle\sum_{l=1}^{N_h}N_{T_k,h_l}
\exp (f(T_k,h_l)-(E_n-h_lM_n)/T_k)} \,,
\end{align}
where $N$, $N_{T_k, h_l}$, $E_n$, and $M_n$ is the total number of data, 
the number of samples associated with $T_k$ and $h_l$, energy of the $n$-th data, and magnetization 
of the $n$-th data, respectively.
This equation should be solved self-consistently for $f(T_i,h_j)$. 
Note that, as in WHAM, $f(T_i,h_j)$ is determined up to a constant.

We repeat the preliminary STM simulations and free energy calculations 
until we finally obtain sufficiently accurate free energy values 
which let the system perform a random walk in the temperature and external field space during the STM simulation. 
We then perform a final production run. 

Note that these two reweighting methods enable us to obtain not only dimensionless free energy values 
but also physical values at any temperature and at any external field.
Such averages are given by 
\begin{align}
\left< A \right> _{T,h} &= \sum_{n=1}^{N} W_{na} A(x_n) \,, \\
W_{na} &= \frac{1}{\left< c_a \right>} \frac{\exp (-(E_n-hM_n)/T)}
            {\displaystyle\sum_{k=1}^{N_T}\displaystyle\sum_{l=1}^{N_h} 
N_{T_k,h_l} \exp (f(T_k,h_l)-(E_n-h_l M_n)/T_k)} \,, \\
        \left< c_a \right> &= \displaystyle\sum_{n=1}^{N} 
            \frac{\exp (-(E_n-hM_n)/T)}
                 {\displaystyle\sum_{k=1}^{N_T}\displaystyle\sum_{l=1}^{N_h} N_{T_k,h_l} \exp (f(T_k,h_l)-(E_n-h_l M_n)/T_k)} \,.
\end{align}
For  details, the reader is referred to Refs.\ \cite{shirts2008statistically,Mitsutake2003}.

We also used two other methods of free energy calculations.
One is given as follows.
By substituting $a(T,h)$ in (\ref{eq:P}) by the estimates for free energy $\tilde{f}(T,h)$, we obtain
\begin{align}
P(T,h) &\propto e^{-f(T,h)+\tilde{f}(T,h)} .
\end{align}
We can write
\begin{align}
f(T,h)=\tilde{f}(T,h) -\ln P(T,h) +\rm{const}.\label{eq:fcal}
\end{align}
Here, $P(T,h)$ can be obtained as a histogram at each set of parameter values in a preliminary STM simulation. 
Thus, this equation enables one to refine the free energy much more easily than the reweighting methods, 
because the method does not require any iterations.
This method does not work well, however,  
when $P(T_i,h_j)$ is too small (or $\tilde{f}(T_i, h_j)$ is  too far away from the true values) 
to obtain samples at $(T_i,h_j)$, while 
the reweighting techniques still work.

The other method for the free energy calculations is a Wang-Landau-like scheme, where we subtract the free energy value being sampled by a fixed constant
during preliminary simulations.  To stand on the safe side, we did not use such data
for reweighting techniques which, strictly speaking, requires equilibrium data as inputs.    
Note that this method also works with inaccurate free energy values. Thus, this method works even when
the free energy estimates are far away from sufficiently accurate values.

In the present work, we first used the reweighting methods and Wang-Landau-like scheme 
to obtain rough estimates of the free energy for the entire parameter space.
We then used the combination of the reweighting methods and (\ref{eq:fcal}) for further refinements of the free energy.

\section{Results and Discussion}
\begin{figure}[bhtp]
   \begin{center}
       \includegraphics[width=5.5cm, clip , angle = 270]{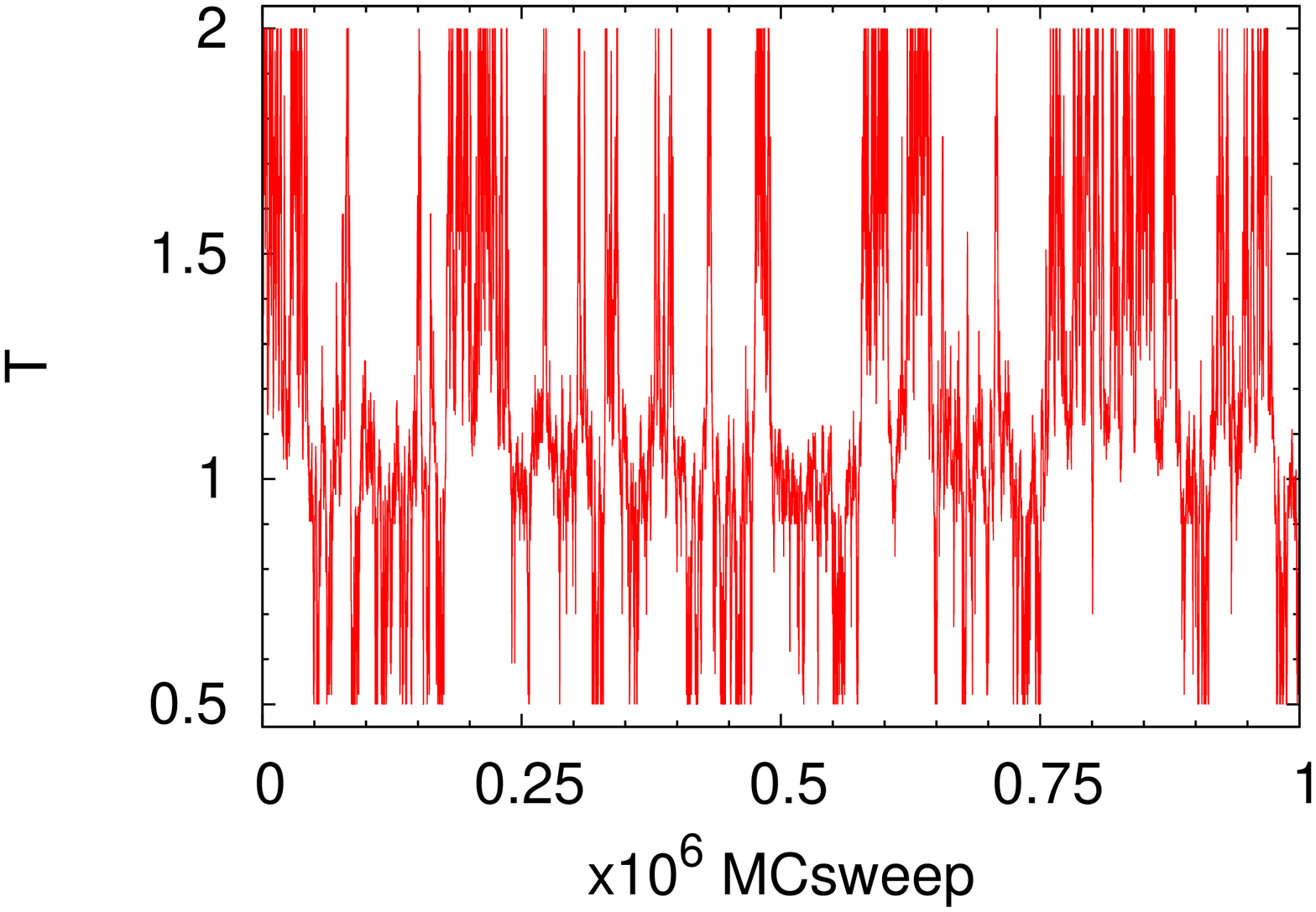}
       \caption{History of temperature $T$ for the linear lattice size $L=80$.}
       \label{fig:temp}
   \end{center}
   \begin{center}
       \includegraphics[width=5.5cm, clip , angle = 270]{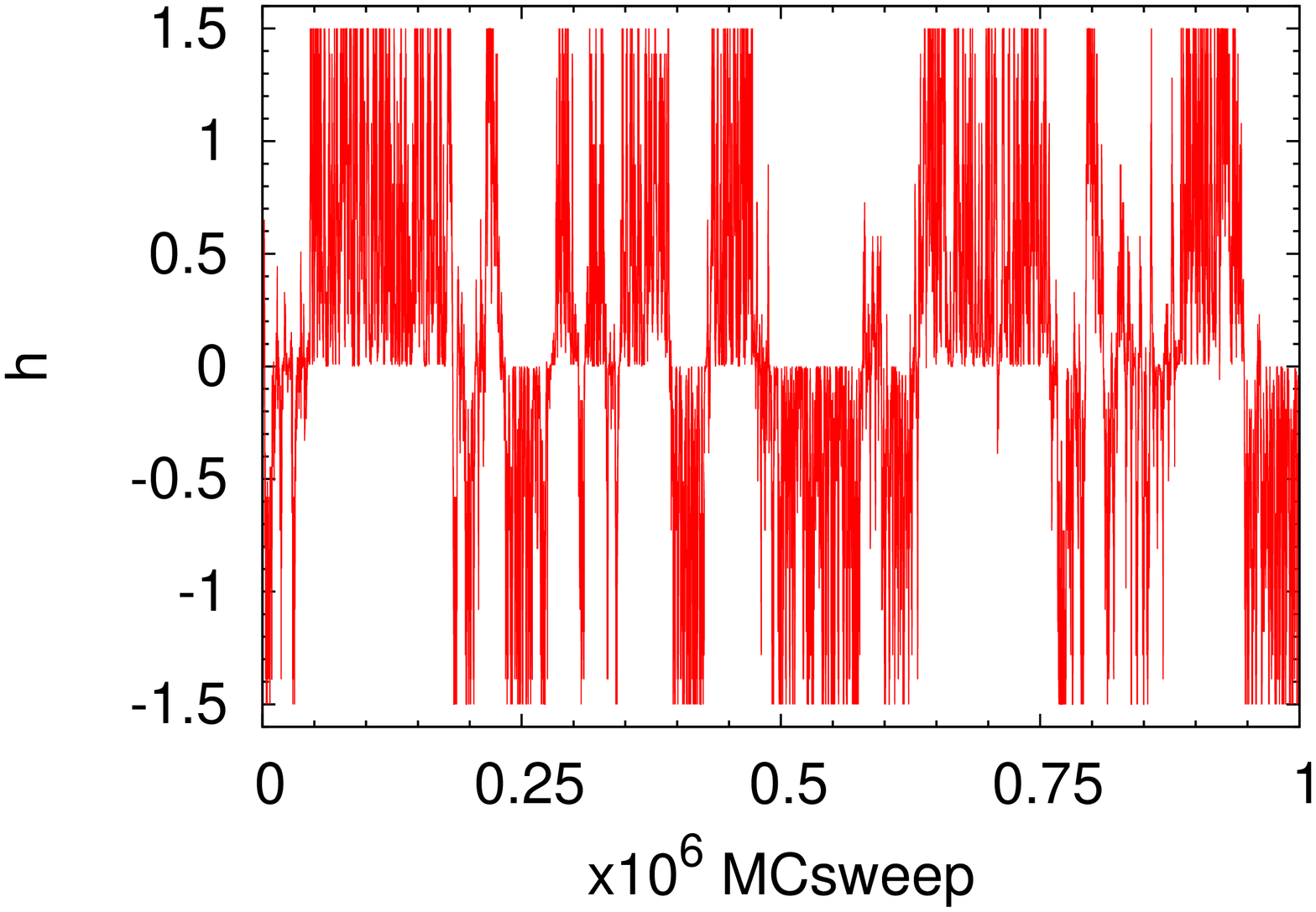}
       \caption{History of external field $h$ for the linear lattice size $L=80$.}
       \label{fig:extf}
   \end{center}
\end{figure}

We  first examine whether the STM simulations were carried out properly or not.
Figure \ref{fig:temp} and figure \ref{fig:extf} show the time series of temperature and external field, 
respectively, for $L=80$. 
In both plots we see block structures reflecting the first-order phase transition line at $h=0$
in the Potts model (see figure \ref{fig:extf}) and the second-order phase transition at the effective Ising transition temperature $T_\mathrm{c}(h)\approx 1.1346$ for
negative external field (see figure \ref{fig:temp}). Within these blocks,  
the temperature and external field indeed realized random walks. 

\begin{figure}[bhtp]
   \begin{center}
       \includegraphics[width=5.5cm, clip , angle = 270]{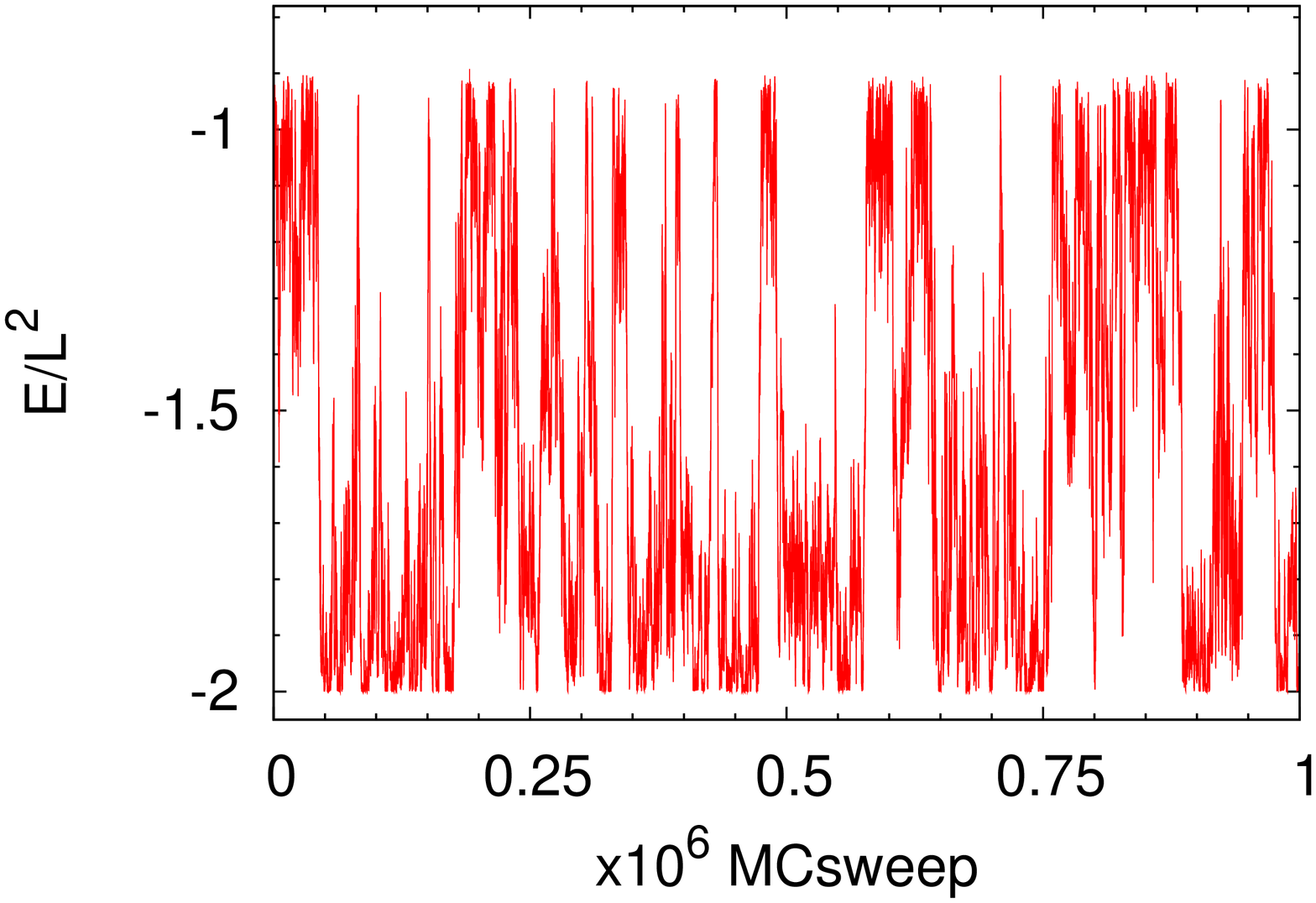}
       \caption{History of energy per spin, $E/L^2$, for the linear lattice size $L=80$.}
       \label{fig:ener}
   \end{center}
   \begin{center}
       \includegraphics[width=5.5cm, clip , angle = 270]{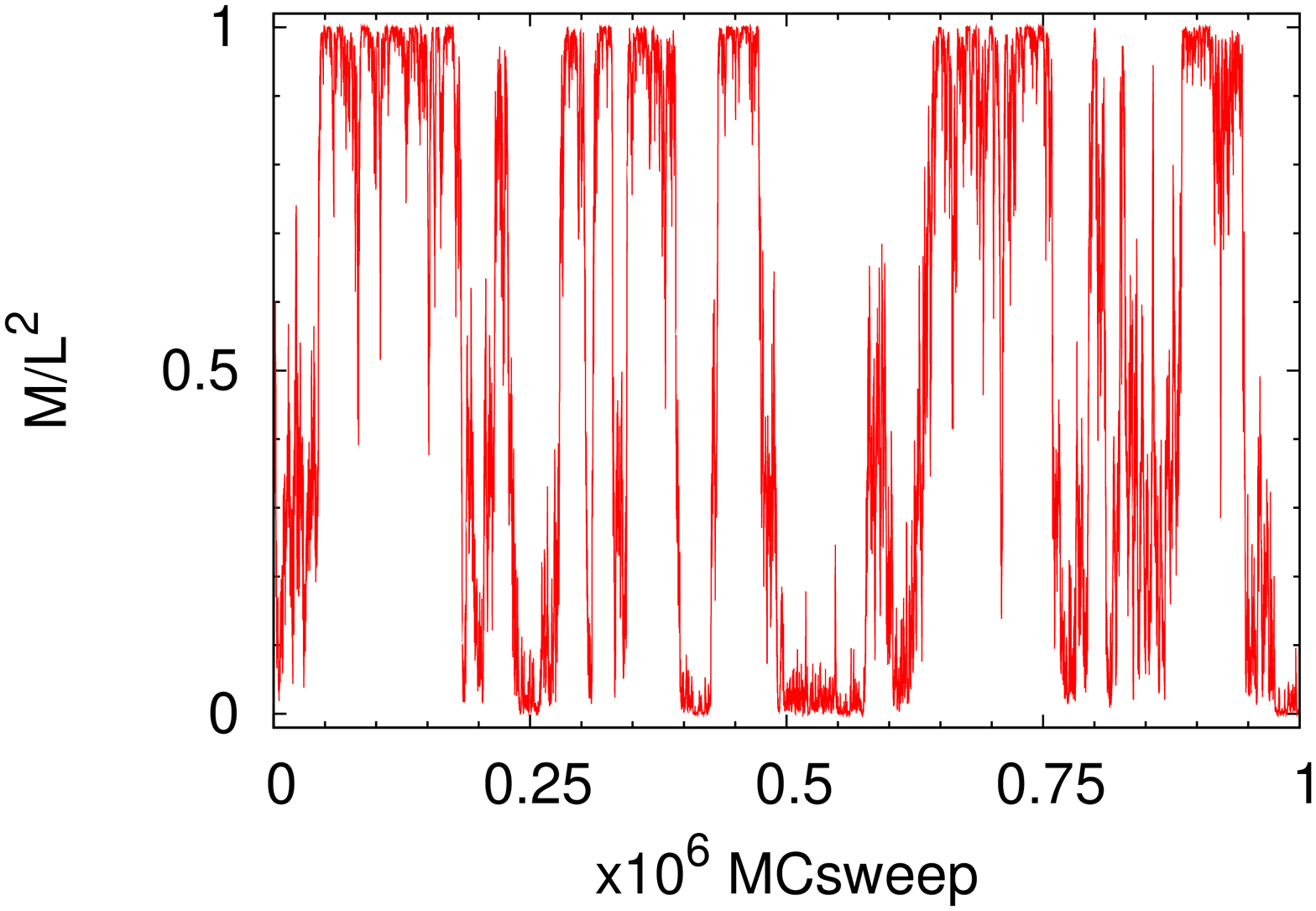}
       \caption{History of the magnetization per spin, $M/L^2$,  for the linear lattice size $L=80$.
        Recall that $M/L^2=1$ corresponds to the ordered state in 0-direction, $M/L^2=0$
		to the ordered states in 1- or 2-direction, and $M/L^2=1/3 $ to the disordered state.  }
       \label{fig:zika}
   \end{center}
\end{figure}

Figure \ref{fig:ener} and figure \ref{fig:zika} show the energy per spin 
and the magnetization per spin, respectively, as functions of MC sweeps. 
They also performed random walks. 
Note that there exist expected correlations between the temperature and energy (see figures\ \ref{fig:temp} and \ref{fig:ener})
and  between the external field and magnetization (see figures\ \ref{fig:extf} and \ref{fig:zika}). 
The same behavior was observed in simulations with  other lattice sizes (data not shown).

\begin{figure}[bhtp]
   \begin{center}
       \includegraphics[width=5.5cm, clip , angle = 270]{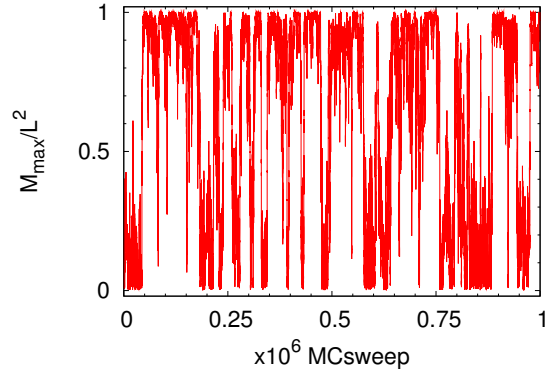}
       \caption{History of $M_{\mathrm{max}}/L^2$ ($ \equiv m_{\mathrm{max}}$), for the linear lattice size $L=80$.}
       \label{fig:zikamax}
   \end{center}
\end{figure}

Figure \ref{fig:zikamax} shows the history of a differently defined magnetization given by 
\begin{eqnarray}
M_{\mathrm{max}} \equiv \left\{ \max_{j=0,1,2} \left[\sum_{i}^{L^2} \delta_{j,\sigma_i}\right] -\frac{L^{2}}{3}\right\} \times\frac{3}{2} .
\end{eqnarray}
Hereafter, we also use the following definition: 
\begin{eqnarray}
m_{\mathrm{max}} \equiv \frac{M_{\mathrm{max}}}{L^2}  .
\end{eqnarray}
This quantity takes physically more intuitive values of 1 and 0 when the system is 
in one of the three ordered phases and in the disordered phase, respectively. 
Here, we see a clear negative correlation between $E/L^2$ and $M_\mathrm{max}/L^2$ (see figures \ref{fig:ener} and \ref{fig:zikamax}).

To compare the results with ordinary ST simulations, we also performed a ST simulation with $L=40$. 
The ST simulation was performed at the conditions similar to those of STM; 
namely, the same total  number of MC sweeps, same temperature distribution, and so on, except that here we set $h=0$. 

\begin{figure}[bhtp]
        \begin{center}
            \includegraphics[width=12cm]{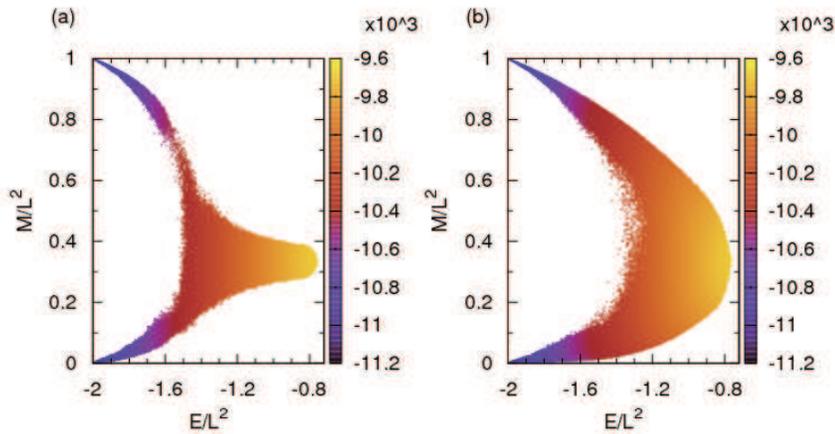}
            \caption{Calculated DOS obtained by WHAM with (a) ST and (b) STM data. 
                      The linear lattice size $L$ was 80.}
            \label{fig:DOS}
        \end{center}
\end{figure}

With the data obtained, we performed  the WHAM calculations to obtain the DOS. 
As shown in figure~\ref{fig:DOS}, the area sampled by STM is larger than that by ST.
Thus, the STM method enables us to perform reweighting techniques in a wider range. 
We recall that $M$ is zero in the 1- or 2-direction ordered phases. The disordered phase corresponds to $M/L^2=\frac{1}{3}$.  

\begin{figure}[bhtp]
        \begin{center}
            \includegraphics[width=9.cm]{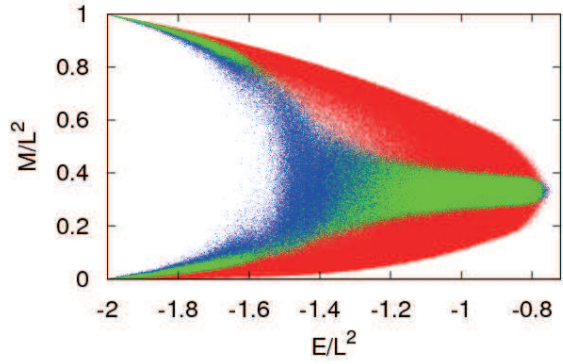}
            \caption{Difference in sampled points between ST and STM. 
                        The red, green, blue, and white regions correspond to the area sampled by only STM, by both of them,  
                         only by ST, and by neither of them, respectively ($L=40$).}
            \label{fig:dosdiff1}
        \end{center}
\end{figure}

We further closely look into the difference in the sampled areas between the two methods. 
Figure \ref{fig:dosdiff1} illustrates how the sampled areas differ.
The red, green, blue, and white regions correspond to the region that was sampled by the STM method exclusively, 
by both methods, by the ST method only,  and by neither of them, respectively. 
Thus, at first sight, it seems that there are some areas in which STM is not good and that ST is somehow 
more powerful than STM. 
\begin{figure}[bhtp]
        \begin{center}
               \includegraphics[width=8.5cm]{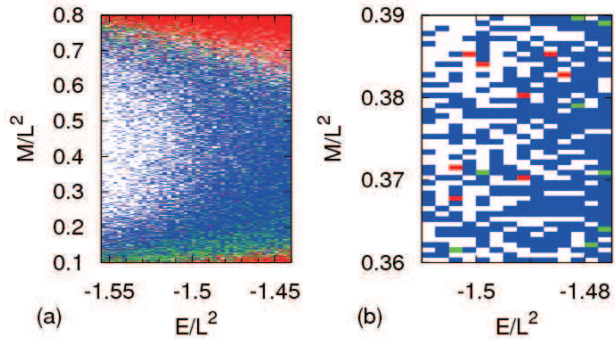}
               \caption{Difference in sampled points between ST and STM. 
                        The red, green, blue, and white regions correspond to the area sampled by only STM, by both of them,   
                         only by ST, and by neither of them, respectively. Plots (a) and (b) are with different focus.}
            \label{fig:dosdiff2}
        \end{center}
\end{figure}

Figure \ref{fig:dosdiff2} zooms in on a region where blue is dominant (mainly sampled by ST). 
There are many pigments (in red and green) which both methods sampled and which even only STM sampled.
This shows that because the ST method has more samples at 
a smaller number of parameter values, the part sampled is narrower but denser. 
However, the representative parts should be sampled properly by STM as well, although the sample density decreases. 

\begin{figure}[bhtp]
   \begin{center}
       \includegraphics[width=5.5cm, clip , angle = 270]{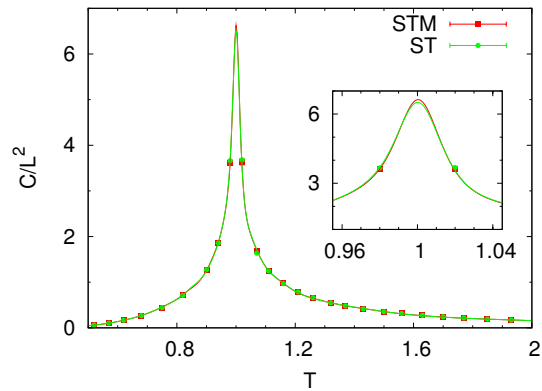}
       \caption{Specific heat $C/L^2$ as a function of  $T$ for $L=40$. 
                    The inset has different abscissa and ordinate. }
       \label{fig:C}
   \end{center}
\end{figure}

\begin{figure}[bhtp]
   \begin{center}
       \includegraphics[width=5.5cm, clip , angle = 270]{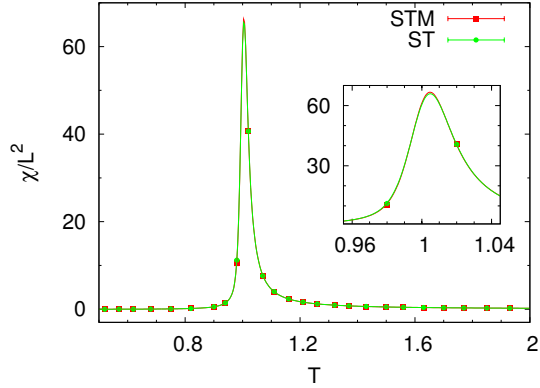}
       \caption{Susceptibility  $\chi /L^2$ as a function of  $T$ for $L=40$.
                    The inset has different abscissa and ordinate. }
       \label{fig:kai}
   \end{center}
\end{figure}

To make it sure that the STM method also samples the relevant area sufficiently, 
we then performed reweighting analyses along $h=0$ with data obtained by ST and STM. 
Figure \ref{fig:C} and figure \ref{fig:kai} show specific heat 
$C/L^2$ and susceptibility  $\chi /L^2$, respectively, as functions of $T$ along the line $h=0$. 
They are defined by
\begin{align}
C &\equiv \frac{\left<E^2\right> - \left<E\right>^2}{T^2} \,,\\ 
\chi &\equiv \frac{\left<M_{\mathrm{max}}^2\right> - \left<M_{\mathrm{max}}\right> ^2}{T}. 
\end{align}

The red and green curves correspond to the data obtained by STM and by ST, respectively. 
The error bars were obtained by the jackknife method \cite{Miller1974, efron1982jackknife, berg2004book, Janke2008}. 
We see  no outstanding differences between the two methods. 
Thus, we confirm that both methods let one sample the representative parts along $h=0$ and that
the STM method enables one to obtain the DOS in wider areas. 

\begin{figure}[bhtp]
   \begin{center}
       \includegraphics[height=6.cm]{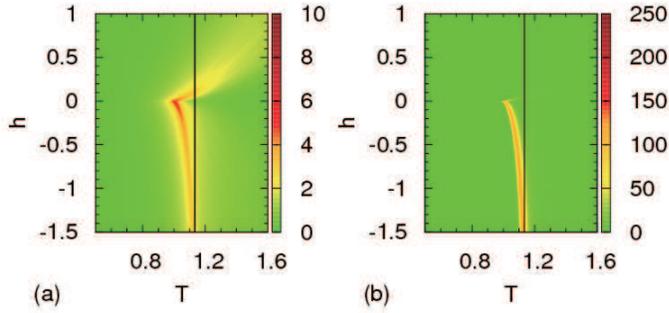}
       \caption{(a) $C/L^2$ and (b) $\chi /L^2$ as functions of  $T$ and $h$ for $L=80$. The solid vertical line 
                              corresponds to $T=1.1346$, which is the critical temperature of the Ising model (in Potts model normalization). }
       \label{fig:Ckaimap}
   \end{center}
\end{figure}

Because the STM method enables us to obtain the DOS in a wide range of sampling space, 
we can calculate the two-dimensional map of any thermodynamic quantity. 
Figure \ref{fig:Ckaimap} shows the specific heat and susceptibility per spin as functions of $T$ and $h$ when $L=80$. 
This implies that the phase transition temperature converges into the Ising case value of $1.1346$, as the external field decreases. 
Related theoretical work is found  in, e.g., Ref. \cite{Sun1991}.

\begin{figure}[bhtp]
    \begin{center}
        \includegraphics[height=8.5cm, clip , angle=270]{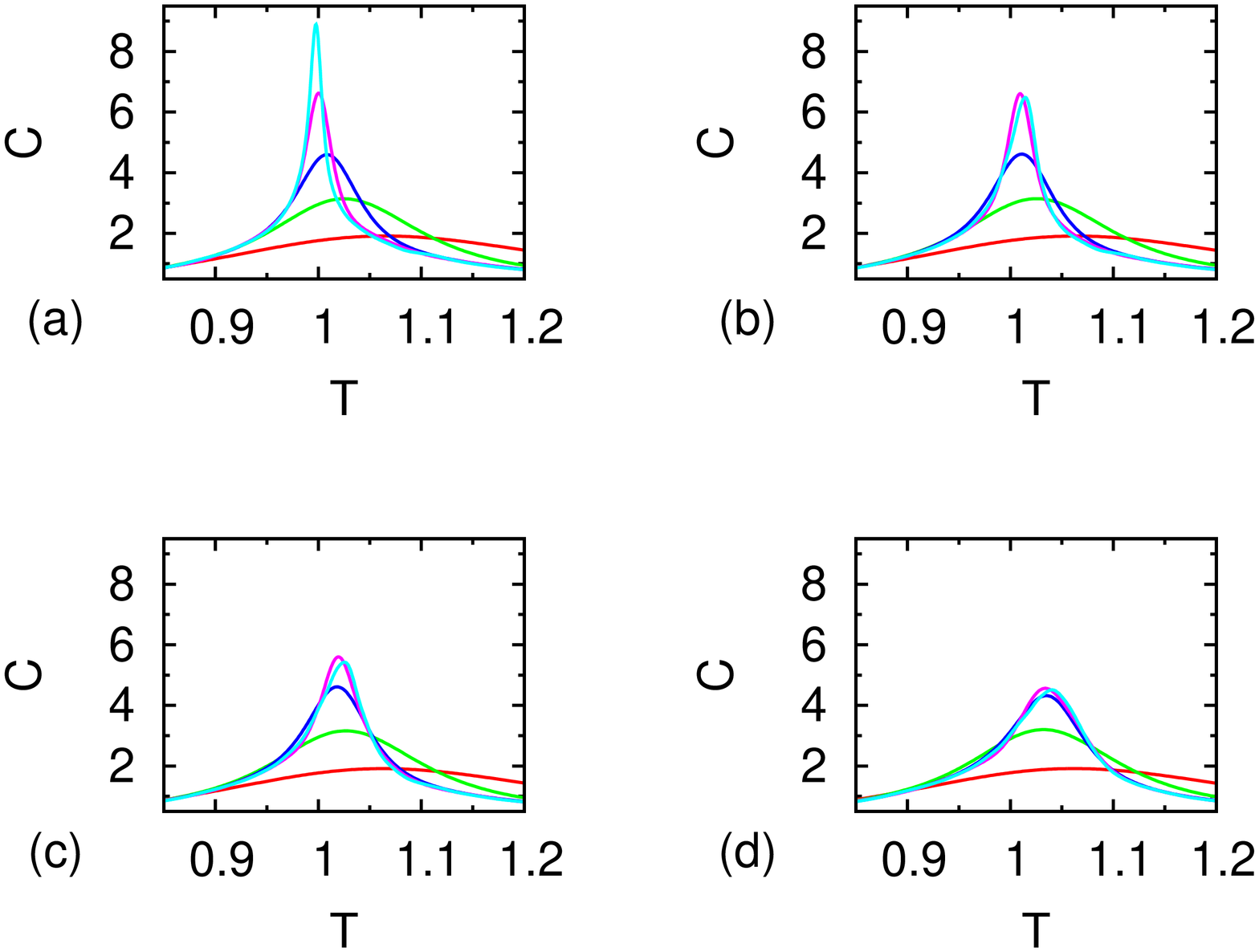}
        \caption{Specific heat $C/L^2$ as a function of $T$ for $L=5$ (red), $10$ (green), $20$ (blue), $40$ (magenta), and $80$ (cyan). 
                     (a) $h=0.0$, (b) $h=0.005$, (c) h=$0.01$ (d) $h=0.02$. }
        \label{fig:C_merged}
    \end{center}
\end{figure}

Figure \ref{fig:C_merged} shows the specific heat as a function of temperature for some values of $h$ and $L$. 
With positive external field, the phase transition disappears. However, because of finite effects, the abnormality,
as measured by the diverging behavior, persists to some extent.  
With smaller external field, the divergence behavior remains for larger $L$. 
Vice versa,  with larger $L$, the more easily it can be shown that the diverging behavior disappears.  
This can be seen as  a crossover between  $L$ and $h$. 

\begin{figure}[bhtp]
    \begin{center}
        \includegraphics[height=8cm, clip , angle=270]{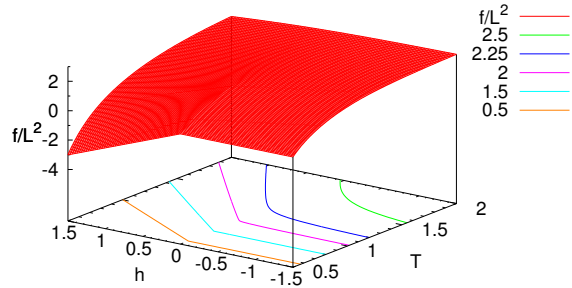}
        \caption{Free energy per spin $f/L^2$ and 
                 its contour curves as a function of $T$ and $h$. The linear lattice size $L$ was 80.}
        \label{fig:fener}
    \end{center}
\end{figure}

Figure \ref{fig:fener} shows the dimensionless free energy per spin as a function of temperature and external field, 
which was obtained by applying MBAR to the results of the production runs.
Note that the partial derivative of this free energy with respect to $h$ 
gives $\frac{\left<M\right>}{TL^2}$, where $\left<M\right>$ is defined in (\ref{eq:M}). 
The shape at $h=0$ suggests a jump of $m$  below $T_\mathrm{c}$, indicating the existence of first-order phase transitions.   

\begin{figure}[bhtp]
    \begin{center}
        \includegraphics[width=12cm]{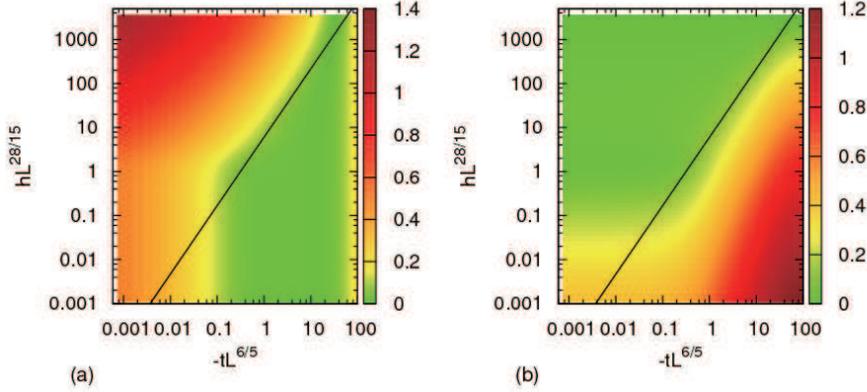}
        \caption{Difference between magnetization and its expected scaling behaviors around the critical point. 
                 The linear lattice size $L$ was 80.
                (a) $|mL^{2/15}-1.2(L^{6/5}t)^{1/9}|$ is illustrated. The solid line corresponds to $h=6t^{14/9}$. 
                (b) $|mL^{2/15}-(L^{28/15}h)^{1/14}|$ is illustrated. 
                 The solid line corresponds to $h=6t^{14/9}$. 
				   }
        \label{fig:3d_scaledMdiff}
    \end{center}
\end{figure}

Finally to study the crossover behavior of the phase transitions, 
we calculated the magnetization by MBAR around the critical point. 
The scaling form of $m$  is given by  \cite{fisher1974renormalization}
\begin{align}
mL^{\beta/\nu}=\Psi (tL^{y_t}, hL^{y_h}),  
\end{align}
where $y_t =1/\nu$ and $y_h=(\beta+\gamma)/\nu$.
According to the crossover scaling formalism  \cite{fisher1974renormalization}, 
if $t^{-\frac{y_h}{y_t}}h$ (in Potts model $t^{-14/9}h$) is small enough, then the magnetization obeys $m\sim t^{1/9}$($=t^{\beta}$), 
and if $h^{-\frac{y_t}{y_h}}t$ (in Potts model  $h^{-9/14}t$) is small enough (i.e., $t^{-14/9}h$ is large enough), then it obeys $m\sim h^{1/14}$ ($=h^{1/\delta}$), 
where $t=\frac{T-T_\mathrm{c}}{T_\mathrm{c}}$. 
Figure \ref{fig:3d_scaledMdiff}(a) shows that if the finite-size effects are negligible 
($L^{6/5}t\gg0.1$) and $t\gg(1/6)h^{9/14}$ 
(i.e., $t^{-14/9}h$ is small), 
then the critical behavior is $m\sim t^{1/9}$. 
Figure \ref{fig:3d_scaledMdiff}(b) shows that if finite-size effects are negligible 
($L^{28/15}h\gg0.1$) and $t\ll(1/6)h^{9/14}$ (i.e., $t^{-14/9}h$ is large), 
then the critical behavior is $m\sim h^{1/14}$. 
Thus, figure\ \ref{fig:3d_scaledMdiff} clearly shows that the line ($t=(1/6)h^{9/14}$) gives 
the boundary of the two scaling regimes. 

\begin{figure}[hbtp]
    \begin{center}
        \includegraphics[height=7.5cm, clip , angle=270]{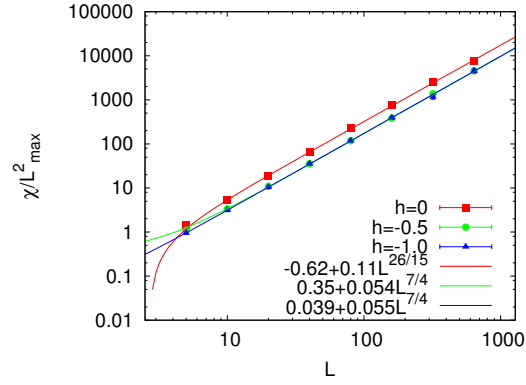}
        \caption{$\chi_{\mathrm{max}}/L^2$ as functions of $L$.}
        \label{fig:kaimax}
    \end{center}
\end{figure}

Because the three-state Potts model in a negative external field is expected to behave like the Ising model,  
we also investigated the crossover behavior between the two models.
The scaling exponents of $\chi_{\mathrm{max}}$ for increasing $L$ of the Potts model and the Ising model are given 
by $\gamma/\nu={26/15}$  and  $7/4$, respectively. 
Figure \ref{fig:kaimax} shows that the exponents are so similar that we cannot distinguish the difference, despite the accuracy of the measurements.
  
Thus, we measured different quantities, which are the maximum values of 
$\frac{\rmd\ln\left<m_\mathrm{max}\right>}{\rmd\beta}$, 
$\frac{\rmd\ln\left<m_\mathrm{max}^2\right>}{\rmd\beta}$, 
$\frac{\rmd\ln\left<U_2\right>}{\rmd\beta}$, 
$\frac{\rmd\ln\left<U_4\right>}{\rmd\beta}$, and 
$\frac{\rmd\left<m_\mathrm{max}\right>}{\rmd\beta}$. 
Here, $U_2=1-\frac{\left<m^2_{\mathrm{max}} \right>}{3\left<m_{\mathrm{max}} \right>^2}$ and 
$U_4=1-\frac{\left<m^4_{\mathrm{max}} \right>}{3\left<m^2_{\mathrm{max}} \right>^2}$ are the Binder cumulants \cite{binder1981finite}.
These quantities were measured by \cite{r355}
\begin{align}\noindent
\frac{\rmd\ln\left<m_\mathrm{max}\right>}{\rmd\beta} &=  \left<E\right>-   \frac{\left<m_\mathrm{max} E\right>}{\left<m_\mathrm{max} \right>}  , \\
\frac{\rmd\ln\left<m_\mathrm{max}^2\right>}{\rmd\beta} &= \left<E\right>-\frac{\left<m^2_\mathrm{max} E\right>}{\left<m^2_\mathrm{max} \right>} ,\\
\frac{\rmd\ln\left<U_2\right>}{\rmd\beta} &=\frac{\left<m_\mathrm{max}^2\right>}{3\left<m_\mathrm{max}\right>^2}
\left\{
\left< E\right> -2\frac{\left<m_\mathrm{max} E\right>}{\left< m_\mathrm{max}\right>}
+ \frac{\left<m^2_\mathrm{max} E\right>}{\left<m^2_\mathrm{max}\right>} 
\right\}, \\
\frac{\rmd\ln\left<U_4\right>}{\rmd\beta}&=\frac{\left<m_\mathrm{max}^4\right>}{3\left<m_\mathrm{max}^2\right>^2}
\left\{
\left< E\right> -2\frac{\left<m^2_\mathrm{max} E\right>}{\left< m^2_\mathrm{max}\right>}
+ \frac{\left<m^4_\mathrm{max} E\right>}{\left<m^4_\mathrm{max}\right>} 
\right\} ,\\
\frac{\rmd\left<m_\mathrm{max}\right>}{\rmd\beta} &= \left<m_\mathrm{max} \right> \left<E\right>- \left<m_\mathrm{max} E\right>.
\end{align}

\begin{figure}[hbtp]
    \begin{center}
        \includegraphics[height=7.5cm, clip , angle=270]{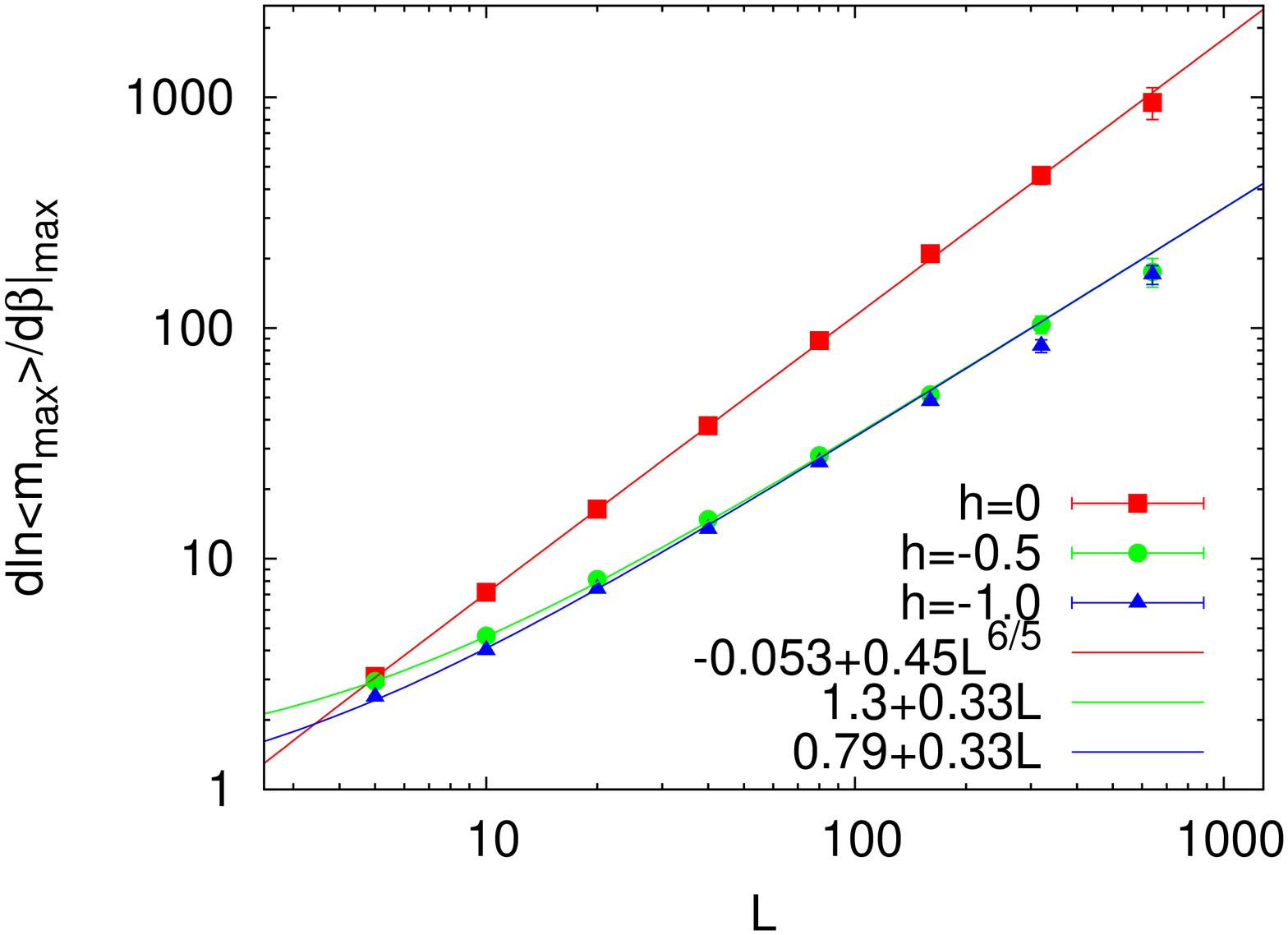}
        \caption{$\frac{\rmd\ln\left<m_\mathrm{max}\right>}{\rmd\beta}|_{\mathrm{max}}$ as functions of $L$.}
        \label{fig:Lcross1}
    \end{center}
\end{figure}

\begin{figure}[hbtp]
    \begin{center}
        \includegraphics[height=7.5cm, clip , angle=270]{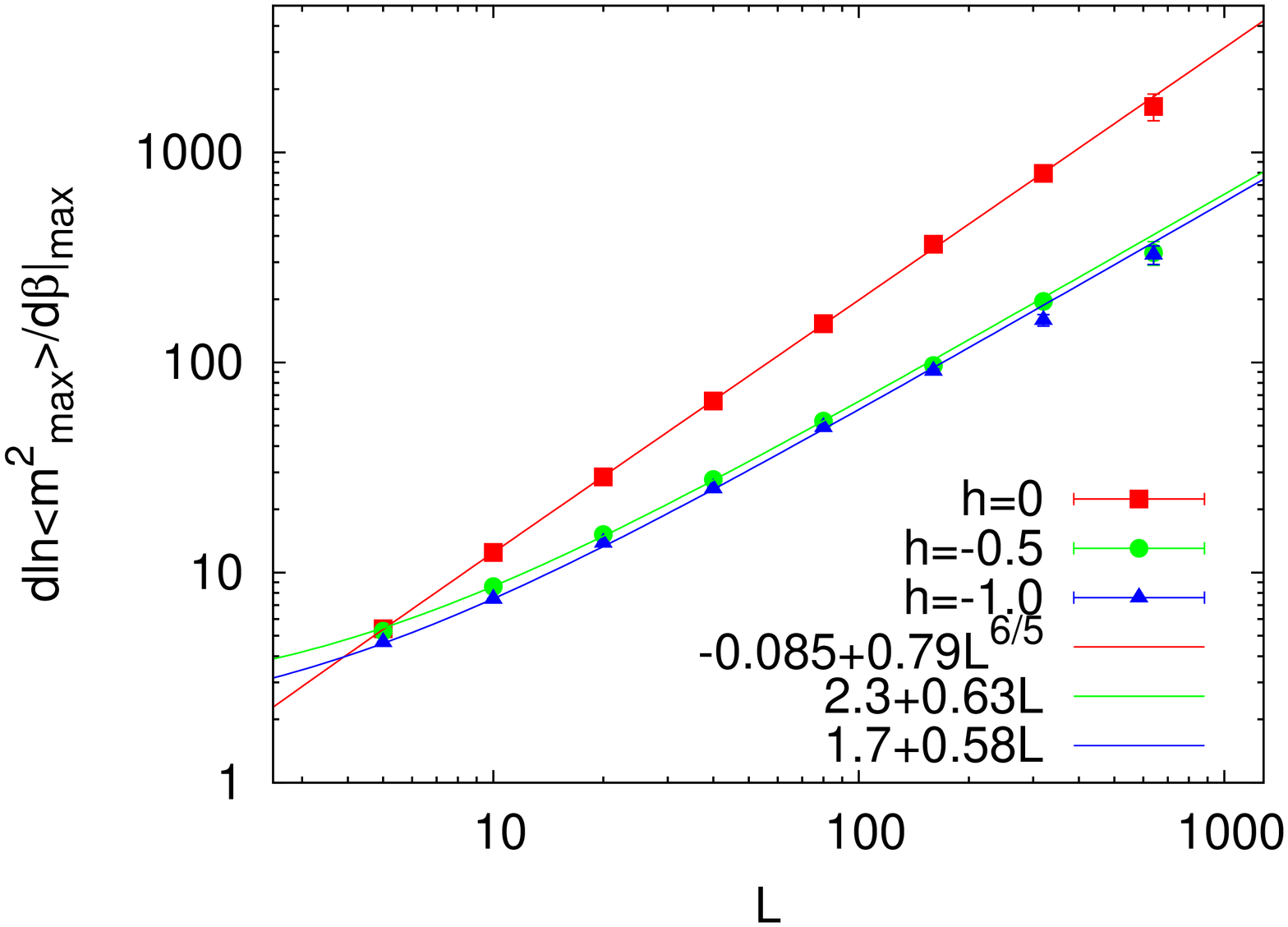}
        \caption{$\frac{\rmd\ln\left<m_\mathrm{max}^2\right>}{\rmd\beta}|_{\mathrm{max}}$ as functions of $L$.}
        \label{fig:Lcross2}
    \end{center}
\end{figure}

\begin{figure}[hbtp]
    \begin{center}
        \includegraphics[height=7.5cm, clip , angle=270]{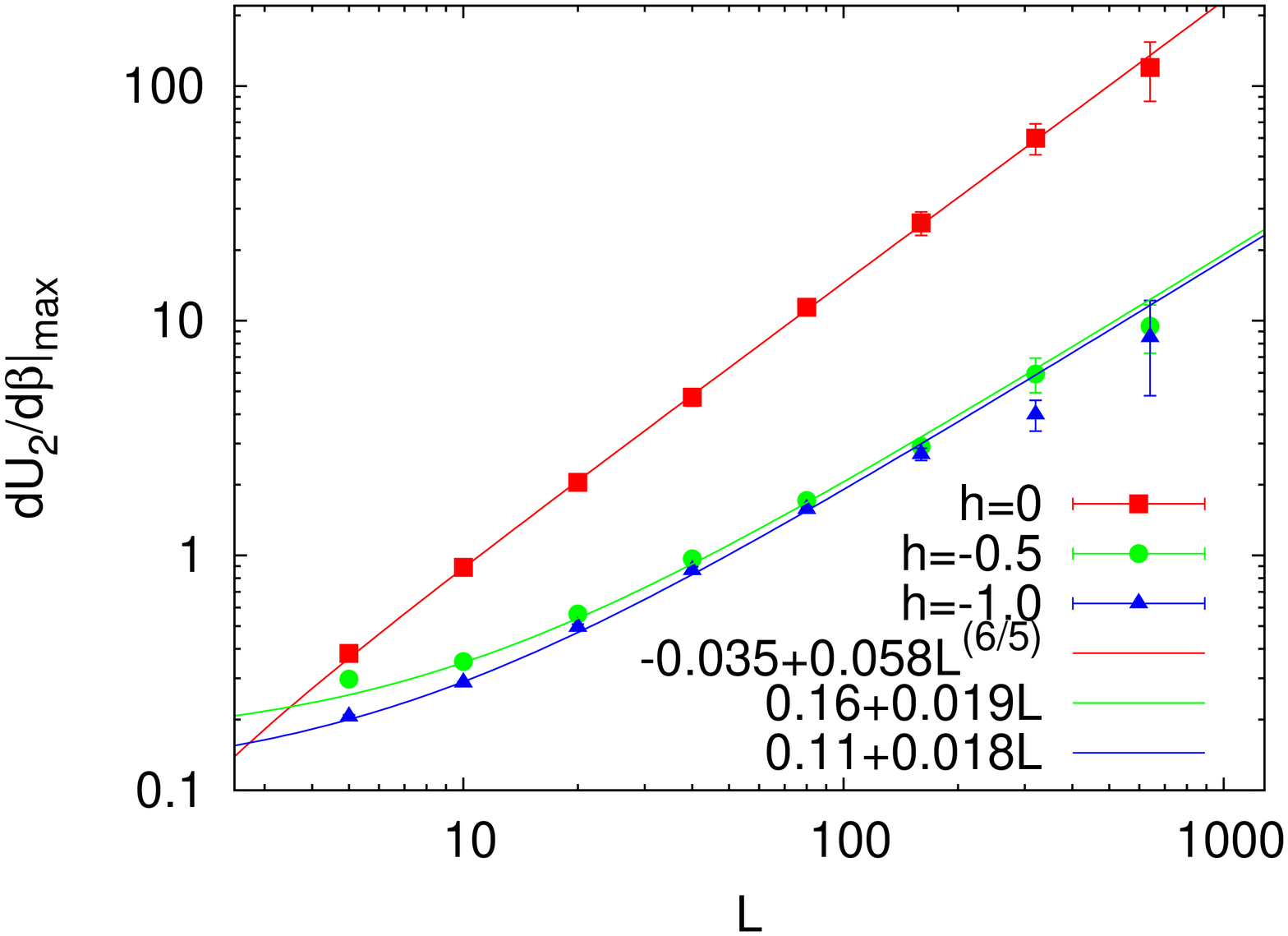}
        \caption{$\frac{\rmd\ln\left<U_2\right>}{\rmd\beta}|_{\mathrm{max}}$ as functions of $L$.}
        \label{fig:Lcross3}
    \end{center}
\end{figure}

\begin{figure}[hbtp]
    \begin{center}
        \includegraphics[height=7.5cm, clip , angle=270]{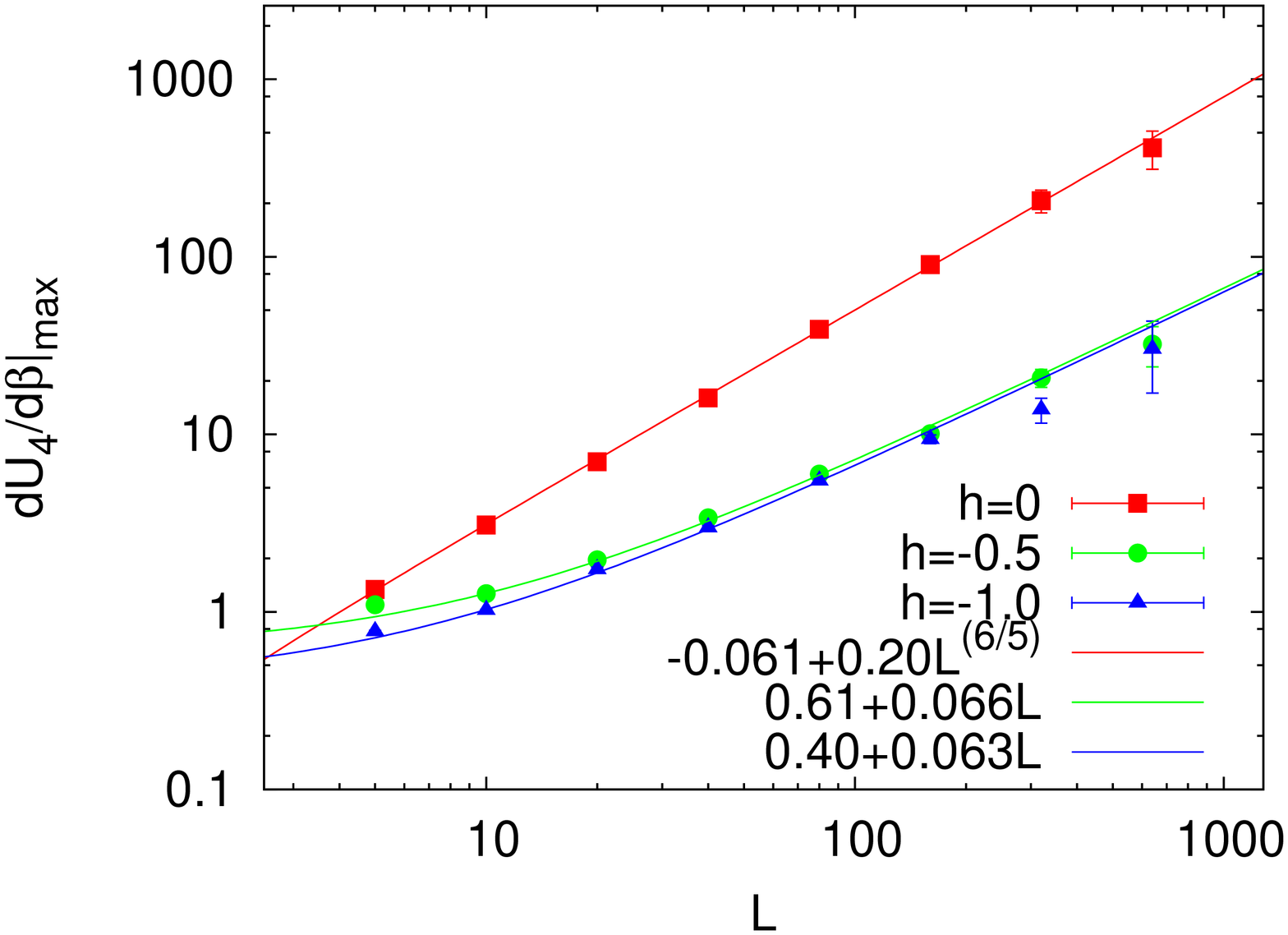}
        \caption{$\frac{\rmd\ln\left<U_4\right>}{\rmd\beta}|_{\mathrm{max}}$  as functions of $L$.}
        \label{fig:Lcross4}
    \end{center}
\end{figure}

\begin{figure}[hbtp]
    \begin{center}
        \includegraphics[height=7.5cm, clip , angle=270]{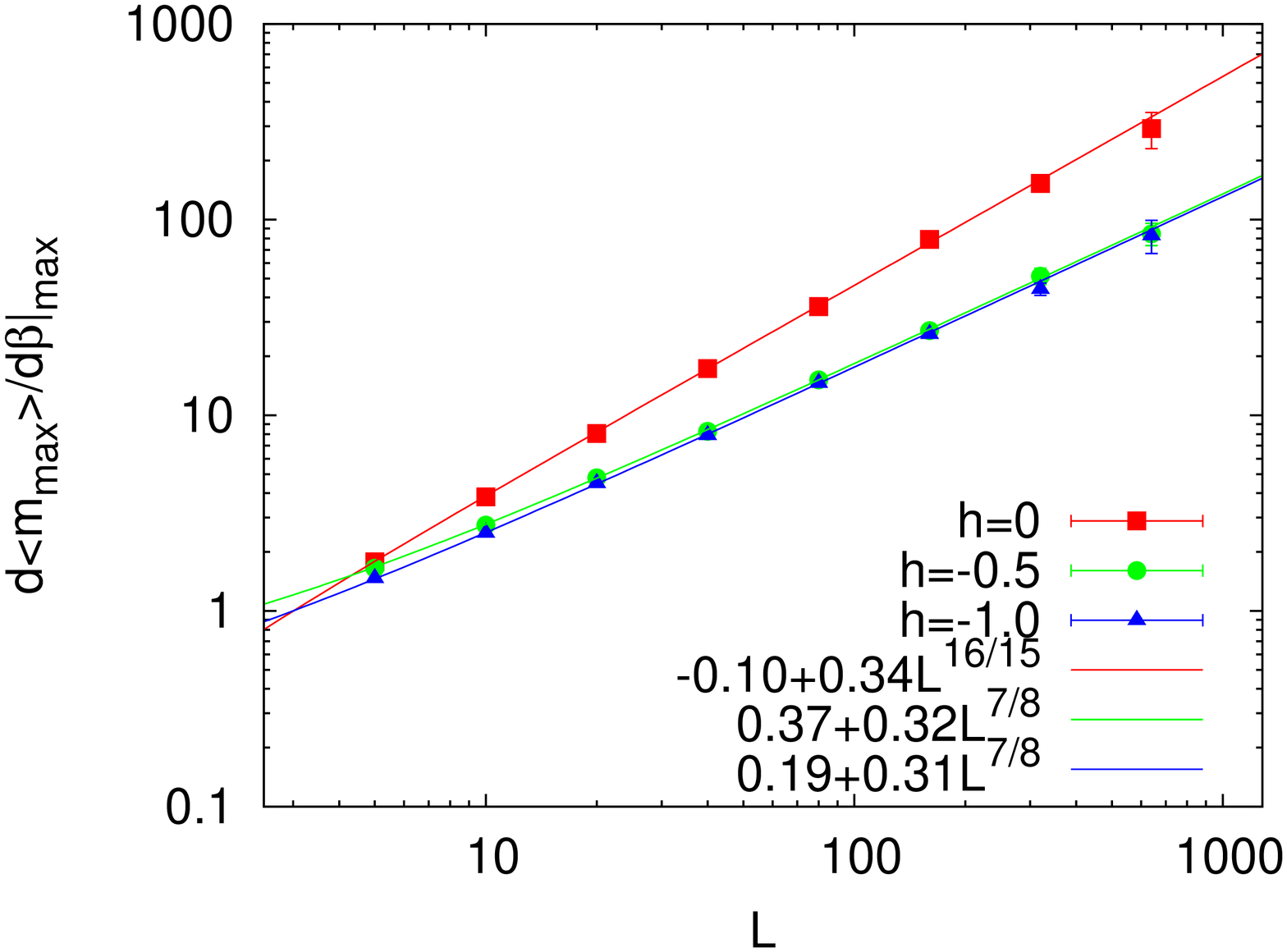}
        \caption{$\frac{\rmd\left<m_\mathrm{max}\right>}{\rmd\beta}|_{\mathrm{max}}$ as functions of $L$.}
        \label{fig:Lcross5}
    \end{center}
\end{figure}

Figures \ref{fig:Lcross1}--\ref{fig:Lcross5} show the results. 
Note that 
$\frac{\rmd\ln\left<m_\mathrm{max}\right>}{\rmd\beta}|_{\mathrm{max}}$, 
$\frac{\rmd\ln\left<m_\mathrm{max}^2\right>}{\rmd\beta}|_{\mathrm{max}}$, 
$\frac{\rmd\ln\left<U_2\right>}{\rmd\beta}|_{\mathrm{max}}$, 
$\frac{\rmd\ln\left<U_4\right>}{\rmd\beta}|_{\mathrm{max}}$, and
$\frac{\rmd\left<m_\mathrm{max}\right>}{\rmd\beta}|_{\mathrm{max}}$
are expected to behave asymptotically as
$L^{1/\nu}$, 
$L^{1/\nu}$, 
$L^{1/\nu}$, 
$L^{1/\nu}$, and
$L^{(1-\beta)/\nu}$, 
respectively, as the lattice size $L$ increases \cite{Janke2008}. 
These critical exponents for  the Potts model  and the two-dimensional Ising model are given by 
 $\nu=\frac{5}{6}$ and $\beta=\frac{1}{9}$ so that $(1-\beta)/\nu =16/15$
and  
$\nu=1$ and $\beta = \frac{1}{8}$ so that $(1-\beta)/\nu = 7/8$ , 
respectively. 
We observe that all quantities along $h=0$ (red curve with filled squares)  follow the Potts case, 
and those with the external field at large $L$ (green curve with filled circles and blue curve with filled triangles) 
follow the Ising case. Note that the two curves at $h=-0.5$ and $h=-1.0$ converge into almost the same line as 
$L$ increases. On the other hand, the curve at $h=-0.5$ (green curve) is more deviating from the scaling behavior. 
This can also be understood as the crossover between $L$ and $h$. 



\section{Conclusions}

In this work, we applied Simulated Tempering and Magnetizing (STM) \cite{Nagai2012Proc, Nagai2012inpress} 
to the two-dimensional three-state Potts model. 
During the simulations, two-dimensional random walks in temperature 
and external field  were realized. 
The random walk covered a wide area of temperature and external field so that the STM simulations enabled us to study
a wide area of the phase diagram from a single simulation run.

Because of the method's capability of  dealing with a wider area of the sampling space (as is seen in DOS), we can calculate 
thermodynamic quantities at an enlarged range of the parameter space. We succeeded in reproducing many typical 
features of the system in the presence of an external field. 

We investigated the crossover behavior of the phase transitions by calculating the magnetization per spin $m$ around 
the critical point by reweighting techniques.      
The results showed agreement with previous theoretical studies. 
Thus, this supports the validity of the two-dimensional ST method, or STM.

With the data of the present work, we can calculate the two-dimensional density of states $n(E,M)$ 
so that we can determine the weight factor for two-dimensional multicanonical
simulations.  Therefore, we can also perform two-dimensional multicanonical simulations 
which would be an interesting future task.

We also remark that the present methods are useful not only for spin systems but also for other complex systems with many degrees of freedom.
Note that because this method does not require one to change the energy calculations, 
the method should be highly compatible with existing program packages.

\ack
Some of the computations were performed on the supercomputers
at the Information Technology Center, Nagoya
University, at the Research Center for Computational
Science, Institute for Molecular Science, and at the Supercomputer Center, 
Institute for Solid State Physics, University of Tokyo. 
This work was supported, in part, by JSPS Institutional Program for Young Researcher Overseas Visit (to T.N.) and 
by Grants-in-Aid for Scientific Research
on Innovative Areas (``Fluctuations and Biological Functions")
and for the Computational
Materials Science Initiative from the Ministry of Education,
Culture, Sports, Science and Technology, Japan
(MEXT). 
W.J.\ gratefully acknowledges support by DFG Sonderforschungs\-bereich SFB/TRR 102
(Project B04) and the Deutsch-Franz\"osische Hochschule (DFH-UFA) under Grant No.~CDFA-02-07.
T.N.\ also thanks the support 
by Nagoya University Program for Leading Graduate Schools: Integrative Graduate Education and Research Program in Green Natural Sciences
for his extended stay in Leipzig.


\section*{References}

\bibliography{citation}

\end{document}